\documentclass[notitlepage]{joli}
\usepackage{times}
\usepackage{pdflscape,pdfpages}

\usepackage{amsmath,amsthm, amssymb}
\usepackage{bm}
\usepackage{widetext}
\usepackage[numbers]{natbib}
\usepackage[normalem]{ulem}

\def\be{\begin{equation}}
\def\ee{\end{equation}}
\def\bea{\begin{eqnarray}}
\def\eea{\end{eqnarray}}

\title[Lunar References Systems, Frames and Time-scales]{ Lunar References Systems, Frames and Time-scales in the context of the ESA Programme Moonlight }
\author*[1,2]{\fnm{Agn\`es} \sur{Fienga}}\email{agnes.fienga@oca.eu, nicolas.rambaux@imcce.fr}
\author[1,2]{\fnm{Nicolas} \sur{Rambaux}}
\author[3]{\fnm{Krzysztof} \sur{Sośnica}}

\affil[1]{\orgdiv{Geoazur}, \orgname{Observatoire de la Côte d'Azur}, \orgaddress{\street{Av. A. Einstein}, \city{Sophia-Antipolis}, \postcode{06560},  \country{France}}}

\affil[2]{\orgdiv{IMCCE}, \orgname{Observatoire de Paris}, \orgaddress{\street{Av. Denfert-Rocheau}, \city{Paris}, \postcode{75014}, \country{France}}}

\affil[3]{\orgdiv{Institute of Geodesy and Geoinformatics}, \orgname{Wrocław University of Environmental and Life Sciences}, \orgaddress{\street{Norwida 25}, \city{Wrocław}, \postcode{50-375}, \country{Poland}}}

\begin{document}


\abstract{
Lunar reference systems represent a fundamental aspect of lunar exploration. This paper presents a review of the topic in the context of the ESA lunar programme, MoonLight. This paper describes the current state of the art in the definition of the lunar reference frame and introduces TCL, a lunar time scale based on IAU resolutions. It also proposes several possible implementations of this time scale for orbiting and ground-based clocks. Finally, it provides an assessment of the improvement of the lunar reference frame that would result from the addition of lunar retro-reflectors on the Moon's surface and the use of orbiter altimetry. This document is an appendix dedicated to lunar reference system definition of a more global document dedicated to the presentation of new concepts in  orbit determination and time synchronization of a lunar radio navigation system.}

\maketitle
\newpage
\tableofcontents
\newpage

In response to the recent need of space agencies to develop new capabilities of robust navigation and positioning infrastructure to support the numerous Moon missions planned for the mid-term future, ESA has established the programme Moonlight  \citep{Giordano21}, for the development of a reliable yet simple lunar communication and navigation service (LCNS).  The fundamental elements of such a system are a precise definition of the reference frames and time scales, accurate satellite ephemerides, synchronisation of time across the constellation and with the ground control centre, stable onboard clocks and the construction of one-way radio signals to be delivered to the end user together with a precise navigation message. 
New concepts for achieving Moonlight goals have been proposed by \citep{Iess2024}, in a work, funded by ESA,  and carried out by a consortium (named ATLAS) that exploited synergies between academic institutions and industries in Europe. 
The present document is the appendix of this work dedicated to lunar reference definition.

\section{State-of-the-art}

\subsection{Introduction}

The definition of a lunar body-fixed coordinate system is essential to
locate a point on its surface and to establish accurate cartography.
These two objectives are particularly crucial in the context of 
lunar exploration. 
Moreover, with the deployment of future lunar constellations, the need of a Moon-centered inertial reference system is also increasing with the intensification and the improvement of the orbit determination of artificial satellites orbiting the Moon.

At present, two slightly different reference systems attached to the Moon are commonly used
to define the lunar body-fixed coordinate system: the Mean
Earth/Rotation Axis (or polar axis) (ME) reference system and the
principal axis (PA) reference system \citep{2018CeMDA.130...22A, williams08}. The former has been used at the beginning of lunar
observation and it is commonly adopted for archiving and data
distribution proposes of lunar surface or topography. The latter
corresponds to the orientation where the lunar tensor of inertia is
diagonal. 
The transformation between the two reference
systems is realised by three static Euler angles that depend on the
lunar gravity field coefficients and dissipative models. Consequently,
this transformation is dependent to a lunar ephemeris. These two
coordinate systems are available in the SPICE kernels in PCK format
containing high-accuracy and the systems are defined through the TF
kernel files \citep{RD32}. Time-varying Euler angles as defined in section
4.9.2 are also provided with the calceph library.

In preparation of the NASA GRAIL mission (2011-2012), three working
groups including the Lunar Geodesy and Cartography Working Group (LGCWG)
\citep{RD32} had proposed to adopt the ME 421 as a reference body-fixed system for
the future scientific data archiving, operations, and communication \citep{RD32}.

Finally, in terms of Moon-centered inertial reference system required for the integration of the orbiter equations of motion, no specific system is yet defined by the community but some have been proposed as, for example, by \cite{2013PhRvD..87b4020T} in the context of the GRAIL mission . They are usually the direct adaptation to the Moon of the GCRS system  defined by the IAU \citep{Soffel2003, 2010ITN....36....1P} . 

\subsection{Inertial and Kinematically non-rotating systems}

Moon-centered inertial reference systems are necessary for the orbit determination of spacecraft orbiting the Moon. Current definitions of inertial systems are based on the IAU B3 recommendations \citep{Soffel2003, 2010ITN....36....1P} and the relativistic descriptions of the BCRS and the GCRS based on \cite{damour1990prd}. The BCRS metric is taken to be kinematically fixed with respect to distant quasi stellar objects (QSO), as defined by \cite{Soffel2003} and following harmonic gauge conditions. 
A kinematically non-rotating eference system centered on the Earth center of mass has been defined in the same type of framework, leading to the definition of the Geocentric Celestial Reference System (GCRS).
Time-scales derived from the BCRS and the GCRS metric are also given in \cite{Soffel2003}.

For the Moon, \citep{2013PhRvD..87b4020T} had proposed to adapt the GCRS definition to the Moon in the context of the GRAIL mission. A description of such a Moon-centered inertial system is given in Sect. \ref{sec:lcrs} with a definition of the Moon centered metric (Sect.  \ref{sec:lcrs_metric})  and time-scale (Sect.  \ref{sec:lcrs_time}). In Sect. \ref{sec:lcrs_acc}, an assessment of the accuracy of two possible realisations of LCRS is given by comparing results obtained with DE421 \citep{folkner2014in} and INPOP19a \citep{2019NSTIM.109.....V}. 

\subsection{Principal axis (PA) system}
\label{sec:PA}

The PA system is defined by the principal axis orientation of the
inertia matrix of the Moon with constant tides contributions. Indeed,
this is a useful reference system to write and to integrate the
rotational equation. Its materialisation is realized from the temporal
series of the Moon orientation, coming from ephemerides, and described
through the rotation Euler angles ($\psi$ precession angle, $\theta$ nutation angle,
and $\phi$ proper rotation). The rotational angles are extracted from the
lunar ephemerides as for example, the JPL ephemerides DE, or the IMCCE/OCA
 INPOP. These two ephemerides are similar in terms of weighted root
mean square residuals. 
Let \emph{I} be a vector in LCRF and PA a vector in the principal-axis reference frame, and \emph{R\textsubscript{x}} and \emph{R\textsubscript{z}} elementary rotational matrices. 
The transformation is written as \citep{folkner2014in}

\begin{equation}
PA\  = \ R_{z}(\varphi)\ R_{x}(\theta)\ R_{z}(\psi)\ I
\label{eq:PA}
\end{equation}

The inverse transformation is:
\begin{equation}
I = R_{z}\left( - \psi \right)R_{x}\left( - \theta \right)R_{z}( - \varphi)PA
\end{equation}

\subsection{Mean Earth direction and rotation system}
\label{sec:ME}
\begin{figure}
\centering
\includegraphics[scale=0.7]{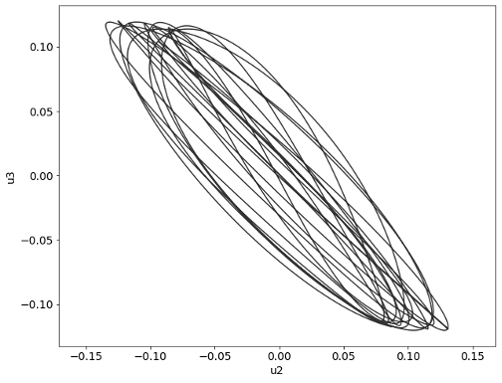}\includegraphics[scale=0.7]{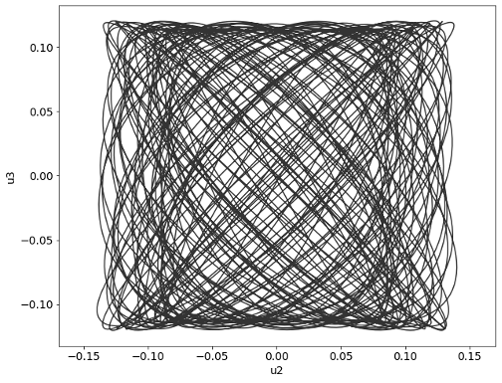}
\caption{Direction to Earth from the Moon for 1 year (as) and 6 years. (u1,u2,u3) are cosine of the direction vector of the Earth seen in PA reference frame
of the Moon. No units. }
\label{fig:ME}
\end{figure}

Due to the fact that the Moon is not purely triaxial ellipsoid, the
average orientation of the PA does not coincide with the mean Earth
direction. The Figure \ref{fig:ME} shows the position of the Earth in the Lunar
PA reference system over 1 and 6 years. The loops describe the
geometrical libration of the Moon. The average of the loops is not
exactly zero and there is a constant offset.

The Mean Earth/Polar Axis (ME) reference system is defined by the
\emph{z}-axis as the mean rotational pole. The Prime Meridian (0$^{\circ}$
Longitude) is defined by the mean Earth direction. The intersection of
the lunar Equator and Prime Meridian occurs at what can be called the
Moon's ``mean sub-Earth point''. This system is an idealization; a
practical attempt to determine these mean directions with high accuracy
would depend on the approach and time interval used. This reference
system is defined from the numerical ephemerides of the PA system
through the following rotation matrix relationship \citep{folkner2014in}
\begin{equation}
ME\  = \ R_{x}( - p_{2c})R_{y}(p_{1c})\ R_{z}( - \tau_{c}\  + \ I_c^{2}\ \frac{\sigma_{c}}{2})\ PA
\end{equation}
where $I_c$ is the Moon moment of inertia, \emph{p\textsubscript{1c}}, \emph{p\textsubscript{2c}}, are
constant coordinates of ecliptic pole in PA reference frame, while
\emph{$\tau$\textsubscript{c}} and \emph{$\sigma$\textsubscript{c}} are constant
libration angles. The offset between these coordinate systems of a point
on the lunar surface is approximately 860 meters and depends on the
ephemeris and on the gravity field associated. Table \ref{tab:euler} reports two
examples with DE and preliminary solution for INPOP.

\begin{table}
\caption{Numerical values of offset angles for some ephemerides. ((a,b)
\citep{williams08},\citep{2011CeMDA.109...85R}, (c) \citep{2021AJ....161..105P}, (d) (Rambaux et al., in progress)).}
\centering
\begin{tabular}{c c c c}
\hline
DE421 (a,b) & -78.513'' & 0.290'' & -67.753''\\
DE440 (c) & -78.694'' & 0.278'' & -67.853''\\
INPOP19a (d) & -78.594'' & 0.290'' & -67.930''\\
\hline
\end{tabular}
\label{tab:euler}
\end{table}

In this case, the inverse transformation is:
\begin{equation}
I = R_{z}\left( - \psi \right)R_{x}\left( - \theta \right)R_{z}\left( - \varphi \right)R_{z}\left( \tau_{c} - I_c^{2}\ \frac{\sigma_{c}}{2} \right)R_{y}\left( - p_{1c} \right)R_{x}\left( + p_{2c} \right)\text{ME}
\end{equation}
\subsection{Semi-analytical series representation}

The IAU WG provided until 2013 the right ascension, declination of the
lunar north pole and the prime meridian from proper rotation angle using
time series. This representation has been stopped in \cite{2018CeMDA.130...22A} because this is less accurate (\textless{} 150 m) than numerical
procedure \citep{2001Icar..150....1K}. Such representation is more compact
than the Chebychev polynomial representation and, if useful, it is
possible to expand the representation to a desire accuracy through
ephemerides such as INPOP.

\subsection{Lunar Laser Ranging Retroreflector as lunar point control}
\label{sec:LRRR}
There are five particular points on the Moon whose positions are very
well measured. They are the
Lunar Laser Ranging Retroreflector (LLRR) and their distribution is
shown in Figure \ref{tab:reflector} and they can be used as lunar control point network.

\begin{figure}
\centering
\includegraphics[scale=0.6]{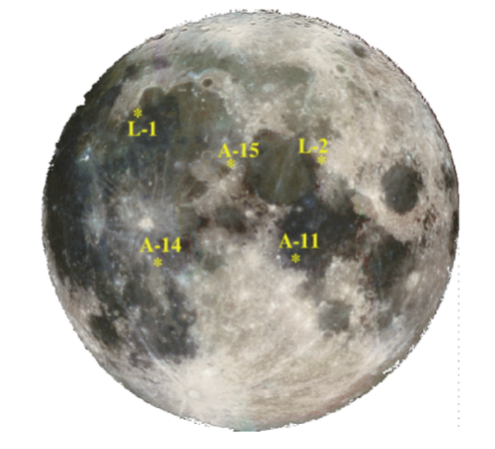}
\caption{Location of the LLRR at the Moon surface.}
\end{figure}

\begin{figure}
\centering
\includegraphics[scale=0.3]{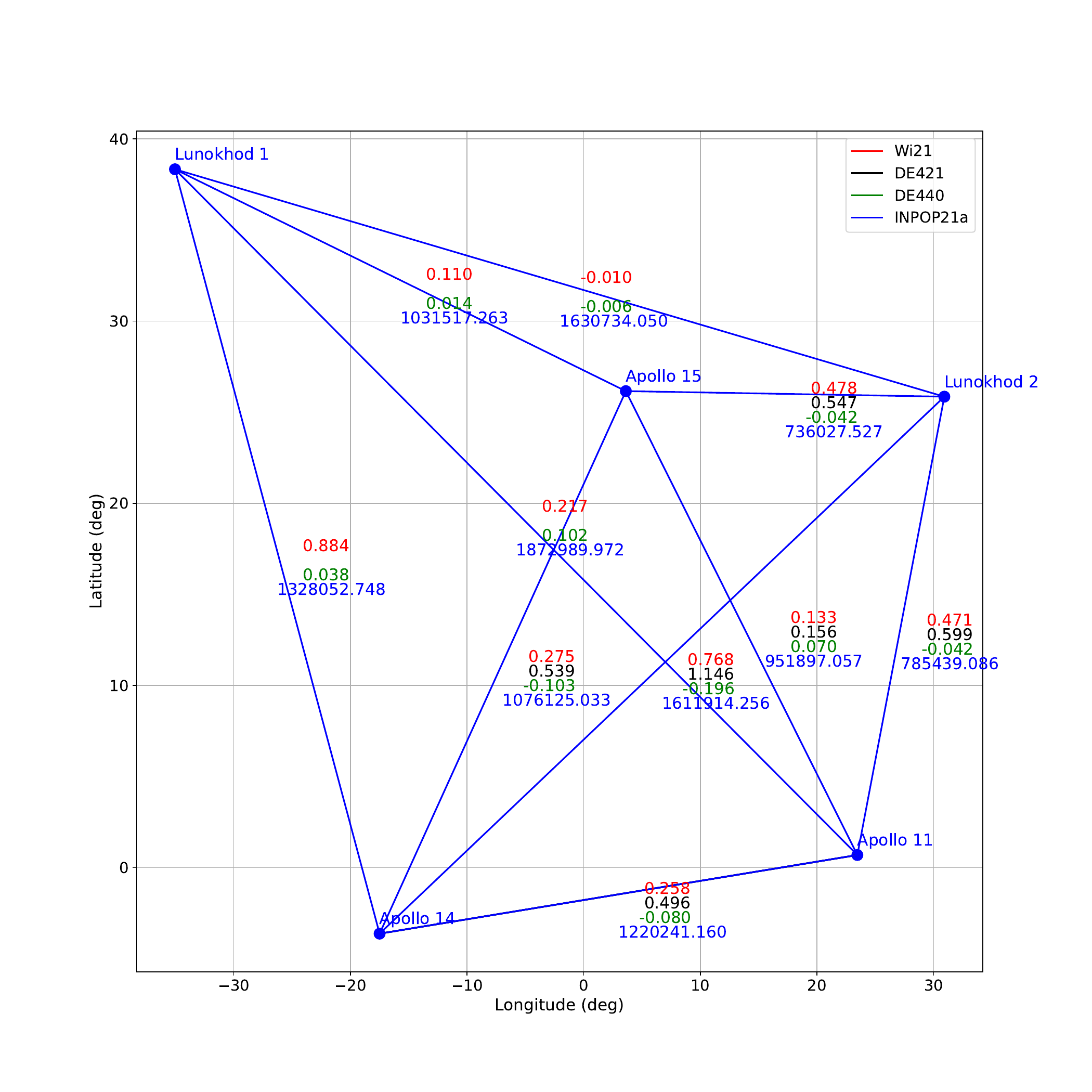}
\caption{Positions of the LLRR and differences in relative positions of
LLRR for INPOP21a (blue value). The other colors represent the
difference in relative positions with respect to the DE ephemerides and
from \cite{2021JGRE..12606920W}.}
\label{fig:posdif}
\end{figure}

The stability of the LLRR positions has been discussed recently in
\citep{2021JGRE..12606920W} who computed the coordinate rates of LLRR.
The weighted average rate of four LLR is 8.3 \(\pm\) 3.1 mm/year, so 20
cm in 25 years. The origin of this shift is not known but they argued
that it could be due to mismodelling in rotational motion. Figure \ref{fig:posdif}
illustrates the differences in positions of the LLRR for different
ephemerides.

Finally, the formal uncertainties in the positions of the LLRR is about 3
cm in INPOP and the global uncertainty in positions is 40 cm for DE \citep{williams08}. These uncertainties of few 10s of centimeters
represent the internal precision in ephemerides. The comparison of
various ephemerides as in Table \ref{tab:reflector} provides the external accuracy that
is about 2 meters for EPM \citep{2020JGeod..94....5P} vs DE \citep{2021AJ....161..105P, folkner2014in}, INPOP \citep{2018MNRAS.tmp...86V, 2019NSTIM.109.....V, 2021NSTIM.110.....F} and 1 meter for INPOP
and DE440.

\begin{table}
\caption{Maximum differences in LLRR positions between INPOP17a \citep{2018MNRAS.tmp...86V}, INPOP19a \citep{2019NSTIM.109.....V}, INPOP21a \citep{2021NSTIM.110.....F}, EPM \citep{2020JGeod..94....5P}, DE421 \citep{folkner2014in}, DE440 \citep{2021AJ....161..105P}.}
\begin{tabular}{c c c c c c}
\hline
LLRR & INPOP17a- & EPM- & INPOP21a- & INPOP21a- & INPOP19a-\\
 & INPOP21a & INPOP17a & DE421 & DE440 & INPOP20a\\
& [m]& [m]& [m]& [m]& [m]\\
\hline
Apollo 11 & 0.108 & 2.3 & 1.361 & 0.898 & 0.078\\
Apollo 14 & 0.101 & 2.2 & 1.601 & 0.838 & 0.051\\
Apollo 15 & 0.148 & 2.2 & 1.409 & 1.030 & 0.091\\
Lunokhod 1 & 0.249 & 2.2 & - & 0.837 & 0.083\\
Lunokhod 2 & 0.186 & 1.9 & 1.156 & 0.949 & 0.061\\
\hline
\label{tab:reflector} 
\end{tabular}
\end{table}

Beside comparing LLRR coordinates obtained in different PA frames as in Table \ref{tab:reflector}, one can also compare these coordinates with those obtained with other technics such as altimetry or imaging. In Sect. \ref{sec:lrs_dtm} are discussed results obtained in using LRO NAC images of the LLRR \citep{10.1007/1345_2015_146} and positions of LLRR obtained in the LRO  Digitized Terrain Model derived by altimetry \citep{glaser19}.


\section{Proposition of lunar reference systems: the LCRS and the LRS}

\subsection{The Moon-centered inertial Lunar Reference system, LCRS}
\label{sec:lcrs}

Based on the concept of the GCRS as given in the IAU 2000 Resolution
B1.3 \citep{Soffel2003,2010ITN....36....1P}, we propose to define an inertial frame, labelled Lunar Celestial
Reference System (LCRS), centered at the Moon center of mass, associated
with a metric tensor based on \cite{damour1990prd} but
adapted to the gravitational environment of the Moon. A lunar time scale
is also defined associated with this metric tensor. This system will be
used for the navigation and the integration of the equations of motion at the
vicinity of the Moon.

\subsubsection{Definition of the metric}\label{definition-of-the-metric}
\label{sec:lcrs_metric}

By analogy with the definition of the GCRS metric tensor and the IAU
2000 Resolution B1.3 \citep{Soffel2003, 2010ITN....36....1P}, we consider the selenocentric metric tensor
S\textsubscript{ab} with selenocentric coordinates (T,
\textbf{X}) with T the Lunar Coordinate Time
(TCL) and \textbf{X}, the coordinate vector of an orbiter in the LCRS. In the same manner, in the BCRS, we note t and \textbf{x}, respectively the barycentric time-scale (TCB or TDB) and the coordinate vector of an orbiter.
The selenocentric tensor is in the same form as the barycentric or the geocentric one but
with Moon-centered potentials L(T, \textbf{X}) and
L\textsuperscript{a}(T, \textbf{X}) such as:

\begin{equation}
S_{00} = \  - 1 + \ \frac{2L}{c^{2}} - \ \frac{2\ L^{2}}{c^{4}}\ ,\ S_{0a} = \  - \ \frac{4}{c^{3\ }}\ L^{a},\ S_{\text{ab}} = \ \delta_{\text{ab}}\left( 1 + \ \frac{2L}{c^{2}} \right).
\end{equation}

The  potentials ${{L}}$ and ${{L^a}}$ should be split into two parts -
potentials \(L_{S}\ \)and \(L_{S}^{a}\) arising from the gravitational action of the Moon and external parts \(L_{\text{ext}}\ \) and
\(L_{\text{ext}}^{a}\ \) due to tidal and inertial effects; the external parts of the metric potentials are assumed to vanish at the selenocenter
and admit an expansion into positive powers of \textbf{X\textsubscript{L}}. Explicitly, one has
\begin{eqnarray}
    {\bf{L}} &=& L_{S} + L_{\text{ext}} + \mathcal{O}\left( {c^{-4}}\  \right) \\
    {\bf{L^a}} &=& L_{S}^{a} + L_{\text{ext}}^{a} = - \frac{G}{2X^3} [ \mathbf{X} \times \mathbf{A_M}] + \mathcal{O}\left( {c^{-2}}\  \right) 
\end{eqnarray}
where $\mathbf{A_M}$ is the Moon angular momentum. $L_S$ correspond to the Newtonian Moon gravitational potential  and $L_{\text{ext}}$ is the tidal contribution produced by all the solar system bodies (excluding the Moon) estimated at the center of mass of the Moon (origin of the LCRS) and at T=TCL. 

For the kinematically non-rotating LCRS, the transformation between BCRS
and LCRS is given explicitly in using \(T\)=TCL, t = TCB, $ \mathbf{r_M} = \mathbf{x} - \mathbf{x_M}$, the
differences between the vectors of the BCRS barycentric position of the orbiter $\mathbf{x}$
and the barycentric position of the Moon $\mathbf{x_M}$, $\mathbf{v_M}$ and $\mathbf{a_M}$ 
being the vectors of its barycentric velocity and
acceleration. The dot stands for the derivative relative to t. 
The expression of the LCRS coordinate vector $\mathbf{X}$ regarding the BCRS $\mathbf{x}$ coordinate vector is given by


\begin{eqnarray}
 & X =  \mathbf{r_M} + \frac{1}{c^{2}} \lbrack  \frac{1}{2} \mathbf{v_M}.(\mathbf{v_M}. \mathbf{r_M}) + l_{\text{ext}}(\mathbf{x_M}).\mathbf{r_M} + \nonumber \\
 & \mathbf{r_M}.(\mathbf{a_M}. \mathbf{r_M}) - \frac{1}{2} \mathbf{a_M}. {r_M}^2 \rbrack + \mathcal{O}\left( c^{- 4} \right),
\label{eq:metricX}
\end{eqnarray}
with $l_{\text{ext}}(\mathbf{x_M}) = \sum_{A\  \neq M}^{}\frac{GM_{A}}{r_{\text{AM}}} + \mathcal{O}\left( c^{- 2} \right)$ with $\mathbf{r_{AM}} = \mathbf{x_A} - \mathbf{x_M}$, the moon-centered vector of the body A in the BCRS.

The time transformation between BCRS time $t$ and LCRS time $T$ is

\begin{eqnarray}
   & T_{L} = t - \ \frac{1}{c^{2}}\ \left\lbrack A\left( t \right) + \ v_{M}^{i}r_{M}^{i} \right\rbrack + \frac{1}{c^{4}}\ \times \nonumber \\
   & \left\lbrack B\left( t \right) + \ B^{i}\left( t \right)r_{M}^{i} + \ B^{\text{ij}}\left( t \right)r_{M}^{j}r_{M}^{i} + C\left( t,\mathbf{x} \right) \right\rbrack + \mathcal{O}\left( c^{- 5} \right),\ 
\label{eq:metricT}
\end{eqnarray}

where $v_{M}^{i}$, $r_{M}^{i}$ and $a_{M}^{i}$ are the coordinates of the vectors $\mathbf{v_M}$, $\mathbf{r_M}$ and $\mathbf{a_M}$ respectively and
\begin{equation}
\frac{\text{d\ A}\left( t \right)}{\text{dt}} = \ \frac{1}{2}\ v_{M}^{2}\  + l_{\text{ext}}\left( \ \mathbf{x_M} \right)\  
\end{equation}
\begin{equation}
\frac{\text{d\ B}\left( t \right)}{\text{dt}} = \  - \frac{1}{8}\ v_{M}^{4}\ {- \frac{3}{2}\ v_{M}^{2}l}_{\text{ext}}\left( \ \mathbf{x_M} \right) + 4\ v_{M}^{i}l_{\text{ext}}^{i}\left( \ \mathbf{x_M} \right) + \ \frac{1}{2}\ l_{\text{ext}}^{2}\left( \ \mathbf{x_M} \right),
\end{equation}
\begin{equation}
 B^{i}(t) = \  - \frac{1}{2}\ v_{M}^{2}\ v_{M}^{i}{- \ 3\ v_{M}^{i}l}_{\text{ext}}\left( \ \mathbf{x_M} \right) + 4\ l_{\text{ext}}^{i}\left( \ \mathbf{x_M} \right),
\end{equation}
\begin{equation}
\ B^{\text{ij}}\left( t \right) = \  - v_{M}^{i}\ \delta_{\text{aj}}{Q^{a} + {2\ \frac{\partial l_{\text{ext}}^{i}}{\partial x^{j}}}\left( \ \mathbf{x_M} \right) - \ \ v_{M}^{i}\frac{\partial}{\partial x^{j}}l}_{\text{ext}}\left( \ \mathbf{x_M} \right) + \frac{1}{2}\text{\ \ }\delta^{\text{ij}}{\dot{l}}_{\text{ext}}\left( \ \mathbf{x_M} \right),
\end{equation}

\begin{equation}
\text{\ C}\left( t,\ \mathbf{x} \right) = \  - \frac{1}{10}\ r_{M}^{2}\ \left( {\dot{a}}_{M}^{i}\ r_{M}^{i} \right),
\end{equation}
\begin{equation}
Q^{a} = \ \delta_{\text{ai}}\ \left\lbrack {\frac{\partial}{\partial x^{i}}l}_{\text{ext}}\left( \ \mathbf{x_M} \right) - \ a_{M}^{i} \right\rbrack
\end{equation}
The expresssion of the potentials $L_{\text{ext}}$, $L_S$ and $L^a$ can be found for example in \cite{2013PhRvD..87b4020T}.
\subsubsection{Definition of the Lunar Time-scale}
\label{sec:lcrs_time}
The UTC time scale is used for the recording of the observations from
the ground. For the definition of the selenodetic frames or when one
uses the Earth-based observations for studying the dynamics of the moon, the UTC
time scale is transformed into TDB \citep{2009AA...507.1675F}.

The GPS Time (GPST) can be used as an intermediate time-scale between lunar orbiters
and ground-based stations as it is proposed in the case of receivers of
Earth GNSS signals in lunar orbit.

For the Moon surface datation or for the constellation orbiting the Moon, we
define TCL, the relativistic time scale at the center of mass of the
Moon, similar to the TCG, the relativistic time scale at the geocenter \citep{Soffel2003,2010ITN....36....1P}. Based on the previous equations, we can defined  TCL as related to  TCB such as TCG is related to TCB following \citep{damour1990prd} with:

\begin{equation}
\frac{\text{d\ TCL}}{\text{d\ TCB}} = 1 + \ \frac{1}{c^{2}}\ \alpha_L + \ \frac{1}{c^{4}}\ \beta_L\ \  + \ \mathcal{O}\left( {c^{-5}}\  \right) \\
\label{eq:TCLvTCB}
\end{equation}

with
\begin{equation}
\alpha_L\  = - \ \frac{1}{2}\ v_{L}^{2} - \ \sum_{A\  \neq L}^{}\frac{GM_{A}}{r_{\text{LA}}}
\label{eq:alpha}
\end{equation}
and

\begin{eqnarray}
\beta_L =& -\frac{1}{8} v_L ^{4} +  \frac{1}{2} \left[ \sum_{A\neq L} \frac{\mu_A}{r_{AL}} \right]^{2} + \sum_{A\neq L} \frac{\mu_A}{r_{AL}} \Bigg\{ 4\bm v_L.\bm v_A - \frac{3}{2} v_L^{2} - 2 v_A^{2}  \nonumber \\
 & + \frac{1}{2}\bm a_A.\bm r_{AL} + \frac{1}{2} \left(\frac{\bm v_A.\bm r_{AL}}{r_{AL}} \right)^{2} + \sum_{B\neq A} \frac{\mu_B}{r_{BA}} \Bigg\},
\label{eq:beta}
\end{eqnarray}

Subscripts A and B enumerate massive bodies, L corresponds to the Moon,
M\textsubscript{A} is the mass of body A, \textbf{r\textsubscript{LA}}=
\textbf{x\textsubscript{L}}-\textbf{x\textsubscript{A}} (vectorial
differences), r\textsubscript{LA} is the norm of the vector
\textbf{r\textsubscript{LA }}, \textbf{x\textsubscript{A }}is the
barycentric vector of position of the massive body A,
\textbf{v\textsubscript{A }}and \textbf{a}\textsubscript{A} are the
barycentric vectors of velocity and acceleration of the massive body A
with respect to TCB. $\alpha L$ is of about 1.48 x 10\textsuperscript{-8}
\citep{2013PhRvD..87b4020T}, and as explained in \cite{Nelson06} and \cite{2013PhRvD..87b4020T},
the term $\beta_L$ Eq. (\ref{eq:beta})  is negligeable for the present applications.

In the same way as TCL can be defined relative to TCB Eq. (\ref{eq:TCLvTCB}), it
is also possible to describe TCL relative to TCG in the GCRS . In this
case, we have
\begin{equation}
\frac{\text{d\ TCL}}{\text{d\ TCG}} = 1 + \ \frac{1}{c^{2}}\ \alpha_{L/G} + \ \frac{1}{c^{4}}\ \beta _{L/G}\ \  + \ \mathcal{O}\left( {c^{-5}}\  \right) \\
\label{eq:TCLvTCG}
\end{equation}
where $\alpha_{L/G}$ and $\beta _{L/G}$ have the same definitions as in Eqs. (\ref{eq:alpha}) and (\ref{eq:beta})  but with geocentric positions and velocities.

\begin{equation}
\alpha_{L/G}\  = - \ \frac{1}{2}\ v_{L/G}^{2} - [ \sum_{A\  \neq L}^{}\frac{GM_{A}}{r_{\text{LA}} } - \sum_{A\  \neq G}^{}\frac{GM_{A}}{r_{\text{GA}} }]
\label{eq:alphaC}
\end{equation}

In the other hand, TT is defined relative to TCG in such a way that \citep{2010ITN....36....1P}
\begin{equation}
\frac{\text{d\ TT}}{\text{d\ TCG}} = 1 - L{_G}
\label{eq:TT}
\end{equation}

with $L_G=6.969290134 \times 10^{-10}$. By considering that TCL-TT=(TCL-TCG)+(TCG-TT), with Eq. (\ref{eq:TCLvTCB}) and the current definition of TT Eq. (\ref{eq:TT}),
it can be deduced that
\begin{equation}
\frac{\text{d\ TCL-TT}}{\text{d\ TT}} = \ \frac{1}{1-L_G}\times [ L_G + \frac{1}{c^{2}} \alpha_{L/C} + \ \frac{1}{c^{4}}\ \beta _{L/C} ] +  \ \mathcal{O}\left( {c^{-5}}\  \right) \\
\label{eq:TCLvTT}
\end{equation}
In Sect. \ref{sec:lrs_time}, are given the results of the integration of Eq. (\ref{eq:TCLvTT}) in the frame of the INPOP planetary ephemerides.

\subsubsection{Transformation between BCRS and LCRS}

From Eqs. (\ref{eq:metricX}) and (\ref{eq:metricT}) and explicit expressions of non-rotating potentials $L$ and $L_a$, the transformation between the local lunar-centered reference frame coordinates of an object in the vicinity of the Moon, \textbf{X(TCL)}, also noted \textbf{X}, and the coordinates of the same object in the BCRS relative to the Moon, \textbf{r$_{M}$(TDB)}, also noted \textbf{$r_M$} as in Eq. \ref{eq:metricX}, is \citep{Soffel2003, 2013PhRvD..87b4020T}
\begin{eqnarray}
 \mathbf{r_M} &= \mathbf{X} + c^{-2} [ - \frac{1}{2} \mathbf{v_M} (\mathbf{v_M} . \mathbf{X}) - \mathbf{X}.l_{ext} + \nonumber \\
& ( \mathbf{w_p}  \times \mathbf{X}) +  \frac{1}{2} \mathbf{a_M}. X^2 - \mathbf{X}(\mathbf{X}.\mathbf{a_M})] + \mathcal{O}\left( {c^{-4}}\  \right)
\label{eq:transfull}
\end{eqnarray}
with $\mathbf{w_p}$ being the vector associated with the relativistic precession (see i.e. \cite{2013PhRvD..87b4020T}), $\mathbf{v_M}$ and $\mathbf{a_M}$ respectively the barycentric velocity and acceleration vectors of the Moon  (same notations as in Sect. \ref{sec:lcrs_metric}), estimated at TDB.  
For the present day application in orbiter dynamics, the Eq. (\ref{eq:transfull}) can be simplified to \citep{2013PhRvD..87b4020T}
\begin{eqnarray}
        \mathbf{r_M}   =  \mathbf{X} [ 1 - c^{-2} \sum_{b \neq M} l_b(\mathbf{r^{b}_M}))] - 
        \frac{1}{2c^2} \left(\mathbf{v_M} . \mathbf{X} \right) \mathbf{v_M} , 
\end{eqnarray}
with $\sum_{b \neq M} l_b(\mathbf{r^{b}_M})$ is the gravitational potential due to the solar system bodies except the Moon, at the Moon center of mass in the BCRS.
For the velocity, with $\mathbf{v}$ is the barycentric velocity of the orbiter in the BCRS at TDB, $\mathbf{V}$, its velocity in the LCRS at TCL, then the transformation for the velocities between BCRS and LCRS is given by \citep{2013PhRvD..87b4020T}
 \begin{eqnarray}
    \mathbf{v} - \mathbf{v_M}   =   [\mathbf{V}  (1 - c^{-2} \sum_{b \neq M} l_b) -   
    \frac{1}{2c^2} \left(\mathbf{v_M} . \mathbf{V}\right) \mathbf{v_M}  ] \times \frac{dt_{TCL}}{dt_{TDB}},
\end{eqnarray}
with $\frac{dt_{TCL}}{dt_{TDB}}$ deduced from by Eq. (\ref{eq:TCLvTCB}).

\subsubsection{Accuracy assessments of the realization}
\label{sec:lcrs_acc}

\begin{figure}
\centering
\includegraphics[scale=0.8]{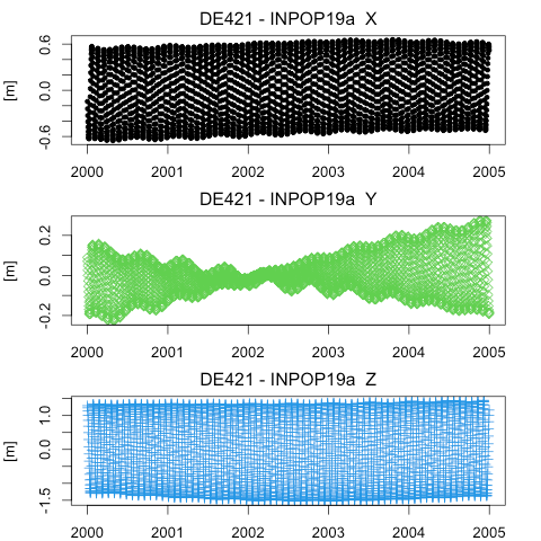}
\caption{Differences in meters on the Moon center positions relative to
the Earth estimated with DE421 and INPOP19a in the ICRS.}
\label{fig:emb} 
\end{figure}

The realization of the LCRS can be done using different lunar and planetary ephemerides. We consider here two generations of ephemerides, DE421
\citep{williams08} which is the present reference for the definition of lunar ME
frame (see Sect \ref{sec:ME}) and INPOP19a \citep{2019NSTIM.109.....V}. These two ephemerides differ as
DE421 model does not account for the core-mantle interactions, the shape
of the fluid core. More than 11 years of additional observations was
also included in the INPOP19a construction. On Fig. \ref{fig:emb} are plotted the
differences in term of localization of the Moon relative to the Earth
using DE421 and INPOP19a over 5 years. The maximum differences of about
1.5 meters are obtained in the Z direction. The ratio between the
gravitational mass of the Earth-Moon barycenter obtained with DE421 and
with INPOP19a differ from unity by $ (1.2 \pm 10^{-11}) \times 10^{-8}  $ .

\subsection{The lunar body-fixed reference system, LRS}
\label{sec:lrs}

In parallel to the LCRS, it is requested to define a frame attached to
the Moon, as the ITRS has been defined as an Earth-attached frame. This
frame, called LRS, will be defined based on the PA properties given in
Sect \ref{sec:PA} The realization of such a system could be obtained  by
using planetary and lunar ephemerides. In the next sections, we consider different accuracy assessments for the LRS realisations. We first consider the differences between the PA realisations using different planetary and lunar ephemerides (Sect. \ref{sec:lrs_acc}). We then consider the propagation  of the PA covariance matrices with time (Sect. \ref{sec:lrs_cov}). We then discuss the
link accuracy with other technics such as altimetry and images obtained with LRO presented in Sect \ref{sec:lrs_dtm}. Finally we consider derived
time-scales that can be more appropriate than the TCL for the user localization or spacecraft navigation based on the definition proposed in Sect. \ref{sec:lcrs_time}.

\begin{figure}
\centering
\includegraphics[scale=0.5]{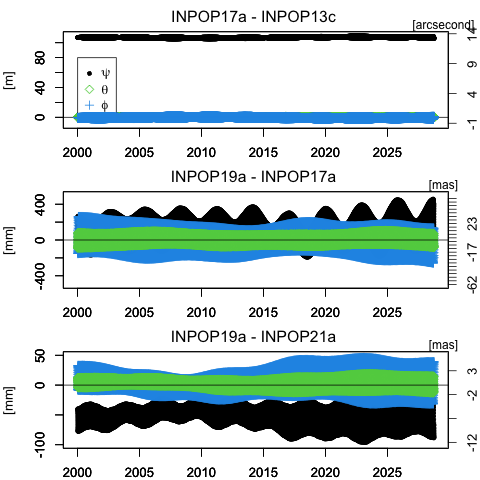}
\caption{Comparisons between the three libration angles of the Moon
obtained with different ephemerides. The plotted differences are given
in terms of displacement on the Moon surface (left-hand side y-axis) and
in milli-seconds of arcs (right-hand side y-axis) and in milli-seconds
of arcs (right-hand side y-axis). }
\label{fig:comparxyz} 
\end{figure}

\subsubsection{Comparisons between ephemerides}
\label{sec:lrs_acc}

At present, the precision and accuracy of LRS realisations are
limited by data accumulation, physical modelling and numerical fit. The
Figure \ref{fig:comparxyz} shows the differences of rotational angles for various
solutions of INPOP. The difference INPOP17a-INPOP13c (black, case I)
represents the improvement in core-mantle interaction and accumulation
of 4 years observations. The difference INPOP19a-INPOP17a (case II) is
related to core-mantle oblateness and 2 years observations and finally
INPOP19a-INPOP21a (case III) is 2 years observations. The improvement in
the physical modeling (case I and II) reaches differences in the order
of 3 meters (with 100 meters offset for \(\psi)\) and 50 centimeters
respectively, meaning that there is still some possibility to improve
the solution at the 50 cm level. The next step for physical modelling is
(not in order of effect): solid core displacement, tidal deformation of
the surface, visco-elastic contribution in deformation coefficients,
impact of Earth Orientation Parameters. The accumulation of data (purely
in case III) improves the solution by 10 cm over 2 years. The principal
impact of observing longer would be to remove some bias introduced in
the reduction analysis, while the principal impact of adding new
reflectors on the Moon surface will be to better disentangle the
different contributions of the libration mis-modeling and to help for a
better study of physical unknowns described previously.

In conclusion, with the most up-to-date modeling, the differences in the
libration angles should not exceed 30 to 50 cm over 5 years.
On Table \ref{tab:comparephem}, we give a summary of the previous comparisons between INPOP solutions but also are indicated the maximum differences in libration angles between INPOP19a and DE421 over the 30 year period.

\begin{table}
\caption{Maximum Differences in libration angles (converted in
meters on the Moon surface) between different generations of ephemerides
over 30 years. Are also indicated the main differences in the dynamical
model and data span between the compared eph}
\begin{tabular}{c c c c c}
\hline
2000-2030 & I17a-I13c & I19a-I17a & I19a-I21a & I19a-DE421\\
\hline
Differences in the model and & CMB interaction & Core shape  &  & CMB interaction, core shape\\
in the data span & + 4 yrs LLR &  + 2 yrs LLR & +2 yrs LLR &  +11 yrs LLR\\
\hline
\(\mathrm{\Delta}\phi\) & 3 m & 40 cm & 6 cm & 3 m\\
\(\mathrm{\Delta}\theta\) & 1.5 m & 15 cm & 3 cm & 50 cm\\
\(\mathrm{\Delta}\psi\) & 3 m & 50 cm & 10 cm & 1.6 cm\\
Offset in  $\psi$ & 100 m & & 5 cm &\\
\hline
\end{tabular}
\label{tab:comparephem}
\end{table}

The main characteristics of INPOP21a can be found in Table \ref{tab:inpopmodele}.

\begin{table}
\caption{Table of interoperability and application to INPOP21a.}
\begin{tabular}{l l l}
\hline
& Parameters/models  & INPOP21a \\
& to be fixed by convention & \\
\hline
LLR data analysis			& & \\
LLR data sample	& by convention with open access	&  \cite{2021NSTIM.110.....F}, Table 4 and Figure 10 \\
station positions	& 	&	Table 7 in \cite{2021NSTIM.110.....F} \\
IRTF to ICRF	& IERS convention 	&	\cite{2010ITN....36....1P} section 5 \\
	
outlier filtering &		& [6] section 5.1.2 \\
bias	& in principle by convention 	&	\cite{vishnu_phd} section 5.1.2 \\
\hline			
Dynamical model			& & \\
Earth tides and gravity field	&IERS convention 2010		& \cite{2018MNRAS.tmp...86V} section 2.2.1  \\
& & \cite{vishnu_phd} section 3 \\
Moon tides and gravity field	&	& \cite{2018MNRAS.tmp...86V} section 2.2.1 \\
& & \cite{vishnu_phd} section 3 \\
Rotational equations		&& \cite{2018MNRAS.tmp...86V}, Eqs 1 and 2a-2e \\
Translationnal equations	&	& \cite{2018MNRAS.tmp...86V} section 2.2.1 \\
Internal structure interactions		& & \cite{2018MNRAS.tmp...86V}, Eqs 1 and 2a-2e \\
& & \cite{2019GeoRL..46.7295V}, Eqs 1-3 and appendix A \\
adjustement weight	& 	& \cite{vishnu_phd} section 5.1.2 \\
method of adjustement	& Least squares  &	\cite{2018MNRAS.tmp...86V} section 2.4 \\
\hline			
LRF definition and delivery			& & \\
Definition of the ME frame	& Eqs in Sec \ref{sec:ME}  & \\
Definition of the PA frame	& Euler angles Eq. \ref{eq:PA}	 & 	 \\
Definition of time-scales	&  Eq. \ref{eq:TCLvTT}	in Sec. \ref{sec:lcrs_time} &  \\
			& & for TT-TDB see Eq. In \cite{2009AA...507.1675F} , Eqs 7 to 10 \\
Delivery of PA frame	& Chebychev polynomials	& calceph \\
&  with fixed orders	& \\
Delivery of time-scales	& depending the precision  & TT-TCL not implemented yet \\
& required & TT-TDB chebychev polynomials  \\
	&  &  based on calceph \\
\hline
\end{tabular}
\label{tab:inpopmodele}
\end{table}

\subsubsection{Propagation of the covariance matrix}
\label{sec:lrs_cov}

An other possible way to assess the accuracy of the LRS realisations is
also to consider its stability according to time. This can be done by
propagating the covariance matrix of the libration angles with time
following the method proposed by \cite{Tapley}. If
H(t\textsubscript{0}) is the covariance matrix of the lunar libration
angles (\(\psi,\ \theta,\ \varphi)\) at the reference time of the
ephemeris t\textsubscript{0} (for INPOP, t\textsubscript{0} is J2000) we
can then study the evolution of the uncertainties H(t) for these angles
at any date t by considering the Jacobian matrix J(t) and with H(t) =
J(t) H(t\textsubscript{0}) J\textsuperscript{t}(t). For this analysis,
we use the covariance matrix and the partial derivatives of INPOP21a.

\begin{figure}
\centering
\includegraphics[scale=0.5]{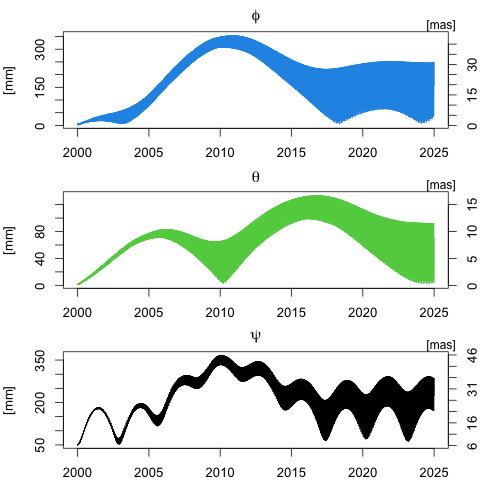}
\caption{Propagation of the INPOP21a
covariance matrix for the three libration angles
(\(\mathbf{\psi,\ \theta,\ \varphi)}\). The results are given in terms
of displacement on the Moon surface (left-hand side y-axis) and in
milli-seconds of arcs (right-hand side y-axis).}
\label{fig:propcov} 
\end{figure}

With the initial uncertainties for (\(\psi,\ \theta,\ \varphi)\) of
about (6.0, 0.1, 0.3) milli-seconds of arcs (mas) respectively, the
result of the propagation of the covariance is given in Figure \ref{fig:propcov}.
The maximum is obtained for \(\psi\) with 35 cm over 25 years. This is
stable and consistent with the comparisons of the recent INPOP versions,
INPOP17a, INPOP19a and INPOP21a.

\subsubsection{Impact of reference frame on gravity field and orbit computations}
\label{sec:gravity}
As it has been explained in Sect. \ref{sec:ME}, the estimations of gravity field from radio-science experiments on board of Moon exploration missions such as GRAIL or LRO have been considering a specific PA frame obtained from the DE421 planetary and lunar ephemerides \citep{2013JGRE..118.1415K}. In the perspective of using different PA frames, it is then important to consider the impact of changing the reference frame in i) the estimation of gravity field coefficients  and ii) in the computation of s/c orbit.\\

\paragraph{Gravity field}

In order to estimate the impact of such a transformation, we first consider the transformation from the GRAIL GL0420A \citep{2013JGRE..118.1415K} gravity field coefficients from DE421 PA frame to LCRF by applying the relation
\begin{equation}
    GF_{LCRF} = R_{z}\left( - \psi_{DE421} \right)R_{x}\left( - \theta_{DE421} \right)R_{z}( - \varphi_{DE421}) GF_{DE421}
\end{equation}
where $(\psi_{DE421},\theta_{DE421}, \varphi_{DE421})$ are the libration angles defining the DE421 PA frame.
Following \cite{2011GeoJI.187..743F}, we also account for the translation from the center of mass of DE421 to the center of mass of INPOP19a at the reference epoch of GL0420A (see Fig. \ref{fig:emb}) and the scaling factor between the Moon mass of DE421 and the one of INPOP19a. The ratio between the mass of the Moon in DE421 and the mass of the Moon in INPOP19a is $1 + 1.2 \times 10^{-8}$.
We then transform the $GF_{LCRF}$ to the $GF_{INPOP19a}$ in using the following relation
\begin{equation}
GF_{INPOP19a}\  = \ R_{z}(\varphi_{INPOP19a})\ R_{x}(\theta_{INPOP19a})\ R_{z}(\psi_{INPOP19a})\ GF_{LCRF}
\end{equation}
with  $(\psi_{INPOP19a},\theta_{INPOP19a}, \varphi_{INPOP19a})$ are the libration angles defining the INPOP19a PA frame.

\begin{figure}
\includegraphics[scale=0.4]{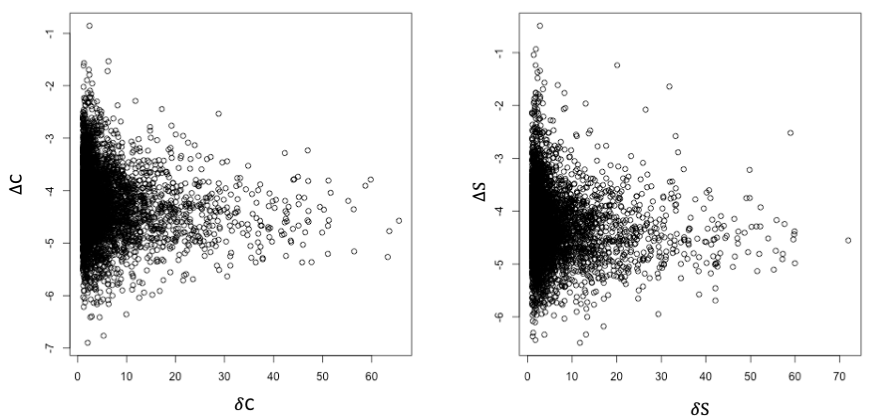}
\caption{Differences between DE421 PA and INPOP19a PA C (on the left) and S (on the right) coefficients over 5 years. The gravity field coefficients are those from GRAIL GL0420A \citep{2013JGRE..118.1415K}. On the x-axis are plotted the ratio between the C and S differences and GRAIL GL0420A uncertainties. On the y-axis, are plotted the log of the ratio of the differences and DE421 C (or S) coefficients as provided by GL0420A. }
\label{fig:GF}
\end{figure}

We note $d_C = C_{DE421} - C_{INPOP19a}$ (respectively $d_S$), the differences in C (respectively S) coefficients in DE421 PA ($C_{DE421}$) and C (respectively S) coefficients in INPOP19a PA ($C_{INPOP19a}$). We then define 
\begin{equation}
\delta C = \frac{d_C}{\sigma_{C}} \, \text{and} \, \Delta C = \frac{d_C}{C_{DE421}}
\end{equation}
where $\sigma_{C}$ the GRAIL GL0420A uncertainty on the C coefficients ($\sigma_{S}$ for S coefficients, respectively). 
On Figure \ref{fig:GF} are plotted $\Delta C$ versus $\delta C$.

The differences were plotted over 5 years. On these plots one can see that the coefficients with differences significantly greater than the uncertainties are those for which the amplitudes are the smallest. In particular, the coefficients have a ratio differences versus uncertainties greater than 10 are those for which the ratio differences versus amplitudes are smaller than 0.01.  

\paragraph{s/c orbit computation}
Differences between lunar reference frames defined on different planetary ephemerides can be organised as follow
\begin{itemize}
    \item Differences in the dynamical scale – the product of the gravity constant and lunar mass. In the case of the DE421 and INPOP19a differences, this difference is of about $1.2 \times 10^{-8}$
    \item Differences in the geometrical scale – the radius of the Moon,
    \item Differences in the orientation of the reference frame – rotation of the selenoid.
\end{itemize}
On the International Centre for Global Earth Models (ICGEM) one can find different lunar gravity field models obtained since 1994. Despite that the latest models are based on GRAIL data, some obvious differences between the models occur. For example, the gravitational mass of the Moon value varies of about $5 \times 10^{-3} \%$ to $2 \times 10^{-3} \%$ between  GrazLGM420b, GRGM660PRIM and  AIUB-GRL350A. Changing  the gravitational mass of the Moon directly impacts the satellite revolution period and satellite radial orbit component. Furthermore, the degree-1 parameters of the lunar gravity field define the origin of the reference frame (the center of mass).
For the gravity field models coinciding with the center-of-mass of a celestial body, the degree-1 coefficients should be nullified. 

To assess the impact of the reference frame differences between DE421 and INPOP19a, the GRAIL gravity field has been re-scaled, rotated, and translated according to the transformation parameters between DE421 and INPOP19a described in Section \ref{sec:gravity}.
We analysed the impact of the reference frame differences on the orbit reconstruction, i.e., the reference orbit has been integrated in the DE421 reference frame and stored as a series of positions (X, Y, Z) every 30 seconds for 1 day. Next, we check the impact of replacing the reference parameters by considering from INPOP19a:
\begin{itemize}
    \item the scale parameter,
\item the translation parameters,
\item the rotation parameters,
\item the scale + translation + rotation parameters, i.e., the full transformation between DE421 and INPOP19a.
\end{itemize}

\begin{figure}
\centering
\includegraphics[scale=0.8]{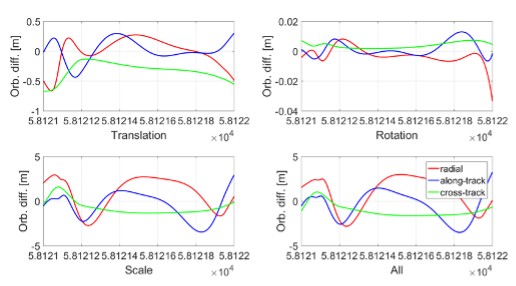}
\caption{Differences between DE421 and INPOP19a due to translation, rotation, scale difference, and all transformation parameters on the orbits derived in one reference frame (DE421) and reconstructed using selected parameters from the second reference frame (INPOP19a)}
\label{fig:k1}
\end{figure}

Figure \ref{fig:k1} shows the differences in the reconstructed orbit of lunar orbiters due to the differences in translation, rotation, scale, and all transformation parameters between  DE421 and INPOP19a. Differences between rotation parameters of a selenoid introduce a marginal effect of -2, 2, and 4 mm of the mean bias for the radial, along-track, and cross-track components, respectively. The RMS error due to the rotation does not exceed 5 mm, thus, can be neglected.  However, the scale and origin must be consistent between reference frames. The mean translation error is equal to 4, 16, and -323 mm for the radial, along-track, and cross-track components, respectively with the RMS values of 238, 179, and 139 mm. The errors generated by the scale differences are at the meter level: 909, -463, and -707 mm for the mean bias and 1775, 1497, and 795 for the RMS of the radial, along-track, and cross-track components, respectively. Thus, we can conclude that the differences of the rotation cause differences at the mm-level, differences of the translation cause differences at the dm-level, whereas differences of the scale cause differences at the m-level for the reconstructed orbits. Therefore, the translation and scale parameters have to be treated consistently, whereas the rotation of the selenoid plays a minor role on the reconstructed orbits. 
The differences between reference frames can be mitigated, to a certain extent, by estimating 7-Helmert transformation parameters that account for the rotation, scale, and origin offset. Table \ref{tab:k1} shows the mean offsets and RMS of the radial, along-track, and cross-track orbit components without and with estimating a posteriori Helmert parameters. The estimated Helmert parameters reduce the mean orbit offsets in the along-track and cross-track components. However large errors remain for the radial (which is sensitive to the dynamical scale difference) and RMS values of all components, because the orbit is integrated and not just geometrically translated, rotated, and re-scaled. Large errors still remain after estimating the Helmert parameters for the integrated orbits; Thus, the consistency in a priori reference frames is indispensable as it cannot be accounted for in the a posteriori transformation.

\begin{table}[]
    \caption{Mean and RMS of differences (in meters) for the three components (Radial, Along-Track and Cross-Track) of the orbits integrated in INPOP19a reference frame w.r.t. DE421 without and with adjusting a posteriori Helmert transformation parameters}
    \centering
    \begin{tabular}{ c | c c c}
    \hline
    & Radial & Along-Track & Cross-track \\
    & m & m & m \\
    \hline
       Without Helmert  & 0.908 $\pm$ 1.846 & -0.447 $\pm$ 1.607 & -1.028 $\pm$ 0.727 \\
       With Helmert  & -0.102 $\pm$ 1.748 & 0.070 $\pm$ 1.484 & 0.00 $\pm$ 0.238 \\
       \hline
    \end{tabular}

    \label{tab:k1}
\end{table}

\begin{figure}
\centering
\includegraphics[scale=0.9]{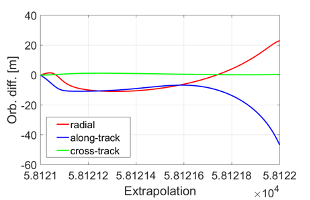}
\caption{Differences between DE421 and INPOP19a on the extrapolated orbit derived as a state vector in one reference frame (DE421) and extrapolated in the second reference frame (INPOP19a)}
\label{fig:k2}
\end{figure}

Finally, we verified the differences between DE421 and INPOP19a for the orbit extrapolation. We took the state vector (position + velocity) for one epoch determined in DE421 and extrapolated the orbit using the numerical integration in the INPOP19a reference frame. This test shows the differences between reference frames in terms of scale, origin, and orientation. 
Figure \ref{fig:k2} shows the impact of the reference frame differences on the orbit extrapolation. After 1-day, the differences exceed 20 and 45 m for the radial and along-track components, respectively. Moreover, as opposed to the orbit reconstruction based on the position series, no RMS values can be estimated when having only the initial state vector, thus, the possible error due to the reference frame differences cannot be identified. 
Therefore, for the orbit predictions, the reference frame must be treated consistently. Information on the satellite state vector must always be accompanied by the information on the reference frame and force models that were the foundation for deriving its value. Otherwise, the errors may reach tens of meters after just one hour of the orbit integration. 

\subsubsection{Link to LRO DTM}
\label{sec:lrs_dtm}

As it has been discussed in Sect. \ref{sec:LRRR}, assessments about the
definition of the lunar frame can be done by considering the
uncertainties of the rotation modelling (see Sect. \ref{sec:lrs_acc} and \ref{sec:lrs_cov} )
but it can also be addressed considering the question of how to link the
lunar reference frame defined in Sect. \ref{sec:LRRR} to the Digital Terrain
Model (DTM) obtained with images and altimetric data from space missions
such as Lunar Reconnaissance Orbiter (LRO). In 2018, a first tie between
LRO DTM and ME DE421 has been obtained by \citep{glaser19} and an other independent tie has been realised with LRO NAC camera \citep{10.1007/1345_2015_146}.
With \citep{10.1007/1345_2015_146}, comparisons between positions of the LLRR observed by LRO NAC cameras and
LOLA altimetric data and positions deduced from JPL DE421 were done and
give maximum differences of about 10 meters, comparable to the RMS of
LRO orbit overlaps \citep{2012JGeod..86..193M}. We have shown in Sect. \ref{sec:LRRR}
 that the differences between LLRR positions deduced from different
lunar ephemerides are of about 1 meter (see Table \ref{tab:reflector}). So, we can infer that the
present link to LRO DTM accuracy is limited by the LRO orbit overlap
accuracy, to 10 meters.

\subsubsection{LRS Time-scales}
\label{sec:lrs_time}
\begin{figure}
\centering
\includegraphics[scale=0.7]{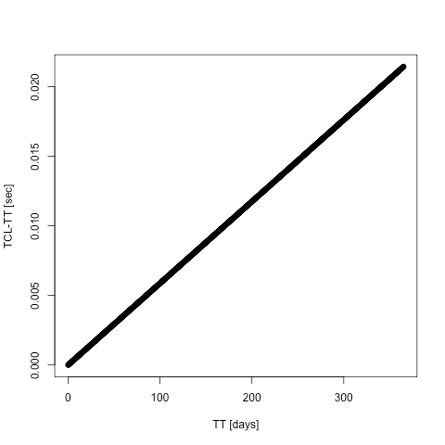}\includegraphics[scale=0.7]{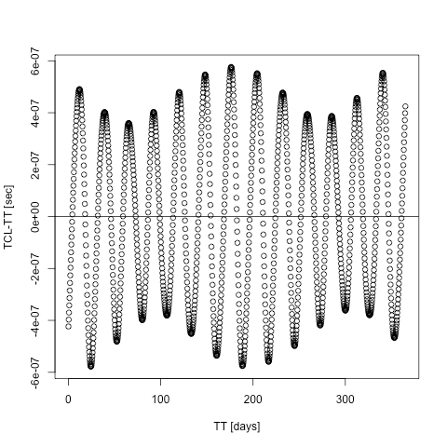}
\caption{On the left-hand side, (TCL-TT) versus Time in TT (days). On
the right-hand side, (TCL-TT) versus Time in TT (days): the remaining
periodic term after the drift removal of about 58. 7\(\mu\)s per day in
(TCL-TT). For the two plots, the results are given in s}
\label{fig:realtime} 
\end{figure}

The time-scales associated with LRS could be a realisation of TCL defined in Sect. \ref{sec:lcrs_time}, but other time-scales can also be proposed depending the requested location of the datation (at the Moon center, at the Moon surface, on board the orbiter).
The Eq. (\ref{eq:TCLvTT}) could be integrated independently from the planetary and lunar ephemeris chosen for the LCRS and LRS realisation using a
simple integrator or directly with the planetary ephemerides. After a
simultaneous integration with planetary orbits, TCL-TT is given in
Figure \ref{fig:realtime}. It appears that TCL-TT can be decomposed in two parts: a
secular drift of about 58.7\(\mu\)s per day and a periodic term with an
amplitude of about 0.6 \(\mu\)s and a period of 27.8 days. These
two terms match with the expected trend that one could obtain by the
addition to \(L_{G}\) of the term accounting for the gravitational
potential of the Earth on the Moon, approximated by --
\(\frac{3}{2c^{2}}\frac{GM_{E}}{a_{M}}\), with a\textsubscript{M}
being the semi-major axis of the geocentric Moon orbit and
GM\textsubscript{E}, the gravitational mass of the Earth. The periodic
term is compatible with the effect of the Moon eccentric orbit and can
be approximated with
\(\frac{2}{c^{2}}\sqrt{GM_{E}a_{M}\ }e_{M}\sin E_{M}\) with
e\textsubscript{M} and E\textsubscript{M} being respectively the
eccentricity and the eccentric anomaly of the Moon orbit. The
differences between the full integration and the approximated terms
(\(L_{G}\ \)--
\(\frac{3}{2c^{2}}\frac{GM_{E}}{a_{M}} - \ \frac{2}{c^{2}}\sqrt{GM_{E}a_{M}\ }e_{M}\sin E_{M}\)
will be of about 0.2 $\mu$s.

The right side of Figure \ref{fig:realtime} shows an additional low frequency
component with a period around \textasciitilde{} 6 months. These other
harmonics are induced by the non-Keplerian motion of the Moon relative
to the Earth. The orbit of the Moon about the Earth is indeed computed
with the full model of perturbations including Sun, planetary
contributions, tides etc... So, it is normal that other harmonics are
present even if we integrate here only the contribution of the Earth
potential on the Moon (perturbed in the full formalism) for the
computation of TCL.

For a datation at the Moon surface, it is necessary to account for the
difference of potential between the time at the selenocenter (TCL) and
the time at the surface (TCLS). TCLS can be obtained for a clock at the
Moon surface with R\textsubscript{c} and V\textsubscript{c} its position
and velocity relative to the selenocenter at time t and
t\textsubscript{0}, such as:

\begin{equation}
    TCL\  - TCLS\  = \ \frac{1}{c^{2}}\frac{GM_{M}}{R_{M}}\ \mathrm{\Delta}t+ \frac{1}{c^{2}}
(R\textsubscript{c}(t).V\textsubscript{c}(t) -
R\textsubscript{c}(t\textsubscript{0})-V\textsubscript{c}(t\textsubscript{0}))
\end{equation}

For a clock at rest on the surface, the second term is 0. The corresponding TCLS is noted TCLS$^{*}$ in the following. 
The first term of the previous equation induces an additional secular drift of about
2.7 \(\mu\)s per day (3.141 x 10\textsuperscript{-11}) for a fixed clock
on the Moon surface. Finally for a Moon orbiter, depending its orbit,
the time on board (TCLO) can be deduced from the TCLS$^{*}$, the time as
indicated by a clock at rest on the surface, by

\begin{equation}
\begin{matrix}
\text{TCLS}^{*}\  - \text{TCLO}\  = \ \frac{2}{c^{2}}\mathbf{R_O\text{~.}} \\
\end{matrix}\ \mathbf{V_O}\  + \left( \ \frac{3}{2c^{2}}\ \frac{GM_{M}}{a} - \ \frac{1}{c^{2}}\frac{GM_{M}}{R_{M}}\  \right)\  \times \ \mathrm{\Delta}t
\end{equation}
with $R_O$ and $V_O$ the position and velocity of the orbiter relative to the selenocenter.
This difference can also be written in supposing that the orbit is elliptic with
\begin{equation}
\begin{matrix}
\text{TCLS}^{*}  - \text{TCLO}\  = \ \frac{2}{c^{2}} \\
\end{matrix}\sqrt{GM_{M}a}\text{\ \ e}\sin E + \ \ \frac{GM_{M}}{c^{2}}\left( \frac{3}{2a}\  - \ \frac{1}{R_{M}} \right) \times \ \mathrm{\Delta}t
\end{equation}
where a is the semi-major axis of the satellite orbit, e its
eccentricity and E its eccentric anomaly.

For a circular orbit with a semi-major axis of 10 000 km and
e=0.01, the effect on the time scale will be a periodical term with an
amplitude of about 1.5 ns with a drift of -2.24 \(\mu\)s per day
(-2.59 x 10\textsuperscript{-11}) when for an eccentric
orbit (e=0.5) with a=5000 km, the amplitude of the periodical effect
will be of about 27 ns with a drift of 1.76 \(\mu\)s per day
(-2.05 x 10\textsuperscript{-11}).

Based on these results, we propose to define a Lunar time scale TL
defined such as
\begin{equation}
 \frac{dTL}{dTCL} = (1 - L_{L}) 
 \end{equation}
 with \(L_{L} = \ 6.796 \times 10^{- 10}\) . 
 The differences between TL
and TT will be then only the periodic part visible in Figure \ref{fig:realtime}.
A broadcast of this periodic term can be proposed as a convenient and
easy way to distribute Lunar Time together with the Ephemeris. Finally,
the correction of the time scale fluctuations induced by the
eccentricity of the satellite orbit (mainly the terms of the TCLS-TCLO)
must be estimated and taken into account in the user's receiver.

The previous results have been obtained in considering only the effect of the Earth on the Moon. If one adds the differential contributions of the Sun and the other planets on the Earth-Moon system, the previous linear trend is still valid, but the period term is modified by a quasi-annual term (1.13 yr) of an amplitude of about 15 µs. The final definition and realization of TCL is currently under discussion.

\section{Simulations for future improvements}
\label{sec:sim}

In this section, we present results of simulations assessing the level
of improvement in the covariance matrix of the libration angles defining
the PA frame (and consequently the LRS). We consider two scenario : in
the first one, we investigate the improvement brought by the
installation of additional retro-reflectors on the Moon surface
following \cite{2022PSJ.....3..136W}; in the second one, we estimate the
improvement brought by altimetric observations obtained from MoonLight
possible orbits.

\subsection{Improvements with additional
reflectors}
\label{sec:sim_lrr}

Through this section, we explore the possibility of an improvement in
the PA angles (see Section \ref{sec:PA}) covariance matrix by having additional reflectors (LRR) on the lunar
surface. In addition to the availability of more LLR (Lunar Laser
Ranging) observations, it also presents an opportunity to have lower
uncertainties in the measurements themselves as both equipment at
stations and analysis models will improve over time. The locations of
the five existing and simulated retroreflectors are listed in Table \ref{tab:coord}.

\begin{table}
\caption{Coordinates of five existing and simulated retroreflector
sites. The stars indicate new tested locations for future possible LLRRs.
}
\begin{tabular}{c c c c c c}
\hline
Site & Long (deg) & Lat (deg) & X (km) & Y (km) & Z (km) \\
\hline
Apollo 11 & 23 & 1 & 1592 & 691 & 21\\
Apollo 14 & -17 & -4 & 1653 & -521 & -110\\
Apollo 15 & 3 & 26 & 1555 & 98 & 765\\
Lunokhod 1 & -35 & 38 & 1114 & -781 & 1076\\
Lunokhod 2 & 31 & 26 & 1339 & 802 & 756\\
South Pole* & 0 & -88 & 61 & 0 & -1736\\
North Pole* & 0 & 88 & 61 & 0 & 1736\\
\hline
\end{tabular}
\label{tab:coord}
\end{table}

LLR measurements can be estimated using a numerically integrated
ephemeris. Several parameters influence accurate range measurements to
LRRs \citep{2018MNRAS.tmp...86V}. These include:

\begin{itemize}
\item
  Initial conditions of the position and the velocity for the Moon (PV)
  with respect to the Earth in the ICRF2 reference frame, at the
  start of the integration time (JD 2451544.5 in TDB).
\item
  Initial conditions of the Euler angles and the angular velocities for
  the full Moon (AL).
\item
  Gravitational mass of the Earth-Moon barycenter
  (GM\textsubscript{EMB}).
\item
  Initial conditions of the angular velocities for the lunar fluid core
  with respect to the mantle frame.
\item
  Polar moment of inertia of the Moon.
\item
  Oblateness of the lunar fluid core which characterizes the difference
  between the equatorial and polar diameters of the fluid core.
\item
  Coefficient of viscous friction at the core-mantle boundary due to the
  relative motion of the lunar mantle and the lunar fluid core at the
  lunar core-mantle boundary.
\item
  Lunar tidal time delay.
\item
  Gravity field coefficients of the Moon.
\item
  Horizontal and vertical tidal lunar Love numbers (h\textsubscript{2,M},
  l\textsubscript{2,M})
\end{itemize}

The covariance matrix with these parameters depicts how each parameter
varies with others and gives the accuracy of the model in comparison to the LLR observations. It is constructed using partial derivatives of each
of these parameters, obtained by simulating ranges and modifying the
ephemeris with appropriate delta (numerical differentiation).

In order to simulate range between ground stations and LLRRs, we use the
ESA software GODOT. GODOT is the ESA/ESOC flight dynamics software for
performing orbit-related computations for estimation, optimization, and
analysis of orbits for mission analysis and in-flight operations
(\url{https://godot.io.esa.int/docs/0.9.0/index.html}). It is intended
as a generic, extensible system for use in practically any space
mission. GODOT can simulate ranges to LLRRs from stations on Earth with
appropriate ephemeris. In order to impose the monthly observational
selection effects for the LLRR sites on the corresponding simulated
ranges, we used existing OCA ground station observations for simulating
ranges. Hence partial derivatives with respect to each parameter and
covariance matrix are obtained following a realistic distribution of the
data. This approach was proposed by \cite{2022PSJ.....3..136W}.

Additional retroreflectors were simulated for times from the beginning
of 2014 till the end of 2021 as per observations to Apollo sites that
were publicly available (\url{http://polac.obspm.fr/llrdatae.html}). We
used existing OCA ground station data for the Apollo 15 reflector for
the times of the simulated observations to the South Pole or the North
Pole in order to impose the monthly observational selection effects for
the Apollo sites on the corresponding simulated data. Since Apollo 15
reflector acquired a lot more ranges than the other sites, we only used
the dates for every third observation \citep{2022PSJ.....3..136W}. \colorbox{white}{Hence 1015
ranges were simulated between 2014 and 2021} to the new additional
retroreflector sites at the South and North poles.

We need to have uncertainty estimates for the observations to the
simulated LLRR. This is accomplished by making four assumptions for each
South Pole and North Pole LLR observation \citep{2022PSJ.....3..136W}.
According to Range Uncertainty Assumption Set 1 or RUAS1, the simulated
range uncertainty is assumed to be the same as that from the OCA Apollo
15 observation. This is a conservative assumption where the ratio of the
uncertainty of the simulated range to the uncertainty of OCA Apollo is 1
even when the new retroreflectors should do better than the older ones.
Whereas in RUAS2, this ratio is 0.5 and is a realistic assumption with
the existing station equipment and analysis model. In RUAS3, the ratio
is 0.1, and equipment at the stations and the analysis model should be
improved to achieve this. In RUAS4, the ratio is 0.01 and this requires
advanced ranging equipment and advanced modeling, which will serve as a
goal for future improvements.

\begin{figure}
\centering
\includegraphics[scale=0.4]{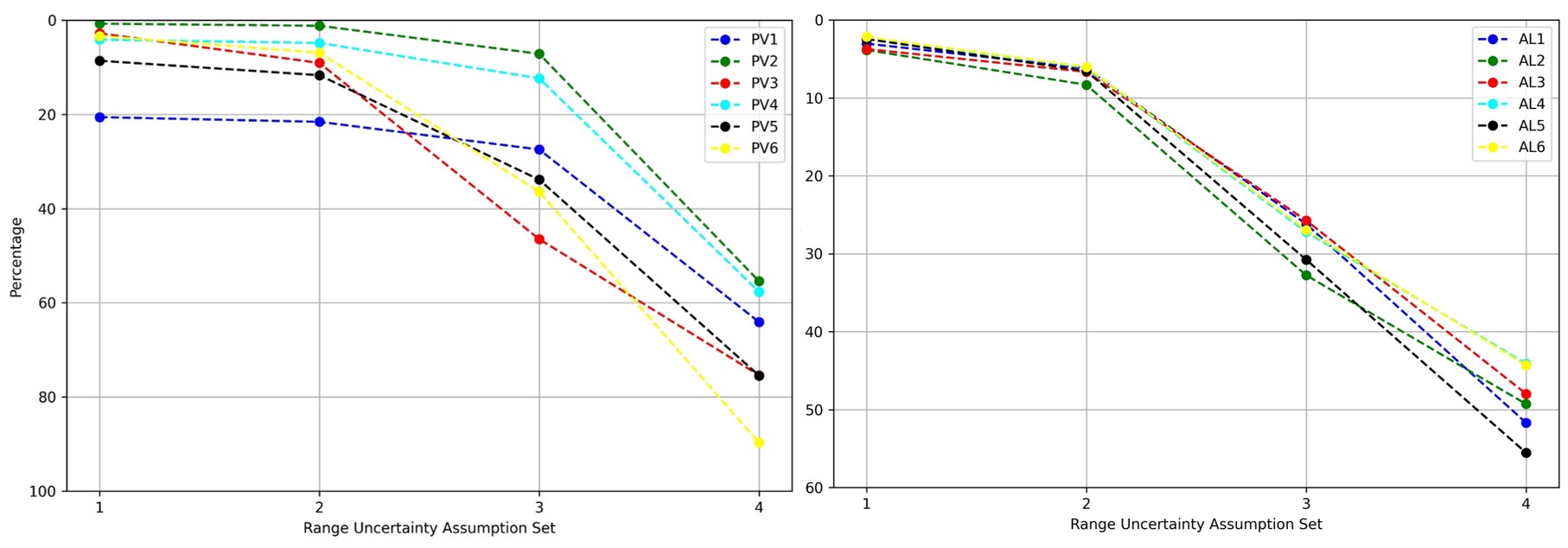}
\caption{ \colorbox{white}{Improvement in \% for positional vector} (PV on the left) and libration
angles (AL on the right) with range uncertainty assumption sets.}
\label{fig:fitPV} 
\end{figure}

The covariance matrix gets updated with the addition of simulated
observations to the new retroreflectors with appropriate RUAS as
discussed above. The uncertainty associated with a parameter, j is given
as,
\begin{equation}
\sigma_j = \sqrt{cov(j,j)}
\end{equation}
where cov are the covariance coefficients. The improvement in percentage, $I_{j}$
in uncertainty of a parameter is calculated as,
\begin{equation}
I_j = \frac{\Delta \sigma_j}{\sigma_j} \times 100
\end{equation}
with $\Delta \sigma_j$, the difference between the uncertainty for parameter $j$ before and
after adding the new LLR observations. 
There will be 4 sets of covariance matrices as there are 4 range
uncertainty assumption sets (RUAS). The greatest improvement in
uncertainties happens for the tidal Love number corresponding to
\colorbox{white}{horizontal solid body tide, l\textsubscript{2} (80\%),} and the gravitational
mass of the Earth-Moon barycenter, GM\textsubscript{EMB} (72\%)
 in the most conservative range uncertainty assumption or RUAS
1. Figure \ref{fig:fitPV} depicts the improvement in positional vector and libration
angles with all four range uncertainty assumption sets.

\begin{figure}
\centering
\includegraphics[scale=0.6]{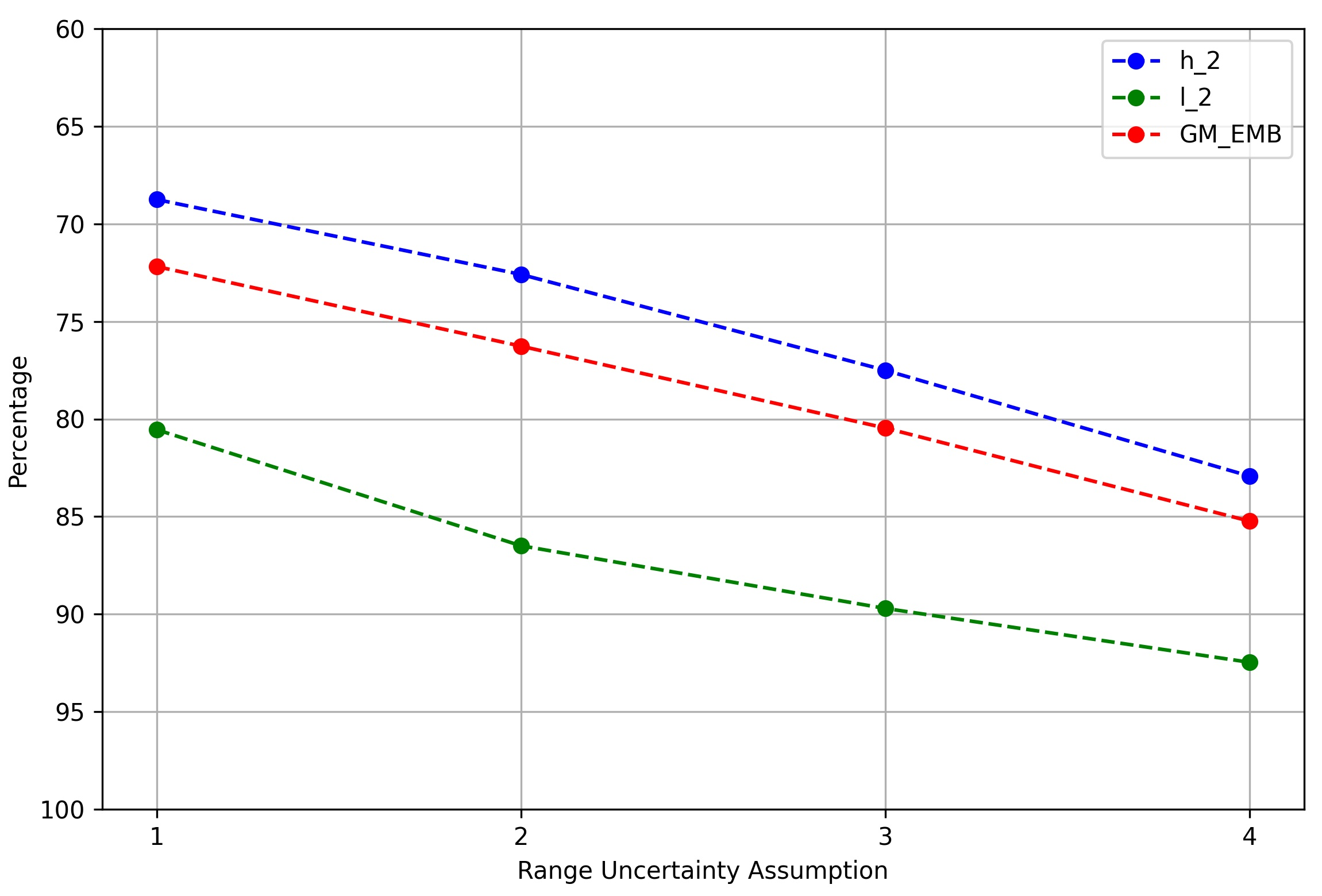}
\caption{ \colorbox{white}{Improvement in \% for h\textsubscript{2}, l\textsubscript{2} and GM\textsubscript{EMB}} with range uncertainty
assumption sets.}
\label{fig:fitemb} 
\end{figure}

For the first RUAS, the improvement in positional vectors (PV) lies
between \colorbox{white}{1\% and 21\%} whereas the addition of retroreflectors causes only
a \colorbox{white}{4\%} improvement for libration angles (AL). But as RUAS goes higher to
RUAS 2, RUAS 3, and RUAS 4, the uncertainty of the simulated
observations goes lower and the weight of these simulations goes higher
in the process of calculating the covariance matrix. In the case of PV,
by RUAS 4 there is improvement in the range of \colorbox{white}{55\% to 90\%} whereas this
value is between \colorbox{white}{44\% and 56\%} for libration angles (AL). Patterns among
components of PV (PV1, PV2, PV3, PV4, PV5, PV6) and AL (AL1, AL2, AL3,
AL4, AL5, AL6) are similar because of high correlation, although not
identical. Improvement in vertical tidal Love number
h\textsubscript{2}, horizontal tidal Love number
l\textsubscript{2} and Gravitational mass of the Earth-Moon barycenter
(GM\textsubscript{EMB}) with range uncertainty assumption sets is shown
in Figure \ref{fig:fitemb}.

In the overly conservative range uncertainty assumption (RUAS 1) gives
an improvement of \colorbox{white}{69\%, 81\% and 72\%} respectively for
h\textsubscript{2}, l\textsubscript{2} and GM\textsubscript{EMB}. This
goes up to \colorbox{white}{73\%, 86\%, and 76\%} for the more realistic RUAS2. It again
goes higher to \colorbox{white}{78\%, 90\%, and 80\%} for RUAS 3 and \colorbox{white}{83\%, 92\%, and 85\%}
for RUAS 4.

Adding new retroreflectors on the lunar surface always adds to the
information available to significantly better define the Earth-Moon system and PA frame definition.

\subsection{Improvements with MoonLight altimetry}
\label{sec:sim_alt}

In addition to the proposed improvements in the definition of the Moon reference system by additional Lunar Laser Ranging retroreflectors (LLRRs), there is a possibility of employing a spacecraft revolving around the Moon equipped with an altimetric device in addition to LRR. The prospective spacecraft \colorbox{white}{('-800')} will carry out laser ranging with the proposed LLRR installed on the Moon’s surface. Based on the present orbital configuration of the MoonLight constellation, it appears that the North Pole LLRR is in this case the most appropriate reflector to be tracked by the laser altimetry on board the spacecraft as it is closest to the currently proposed spacecraft orbit.  The initial conditions (2026 June 01 00:00:00 TDB) for the spacecraft orbit from the center of the Moon in the J2000 frame are given in Table \ref{tab:cdi}.

\begin{table}
\centering
\caption{Moon-centered spacecraft cartesian initial conditions given at 2026 June 01 00:00:00 TDB}
\label{tab:cdi}
\begin{tabular}{c c c }
& Positions & Velocities \\
& km & km.s$^{-1}$ \\
\hline
$x$ & -2181.20  & -0.32 \\
$y$ & -9576.67   & -1.29\\
$z$ &  2601.48     & -0.72\\
\hline
\end{tabular}
\end{table}

The scope of these simulations is not to discuss the feasibility of such measurements but  to assess the gain for the definition of the PA frame if such altimetric campaign takes place with different range of possible accuracies.

\begin{figure}
\centering
\includegraphics[scale=0.45]{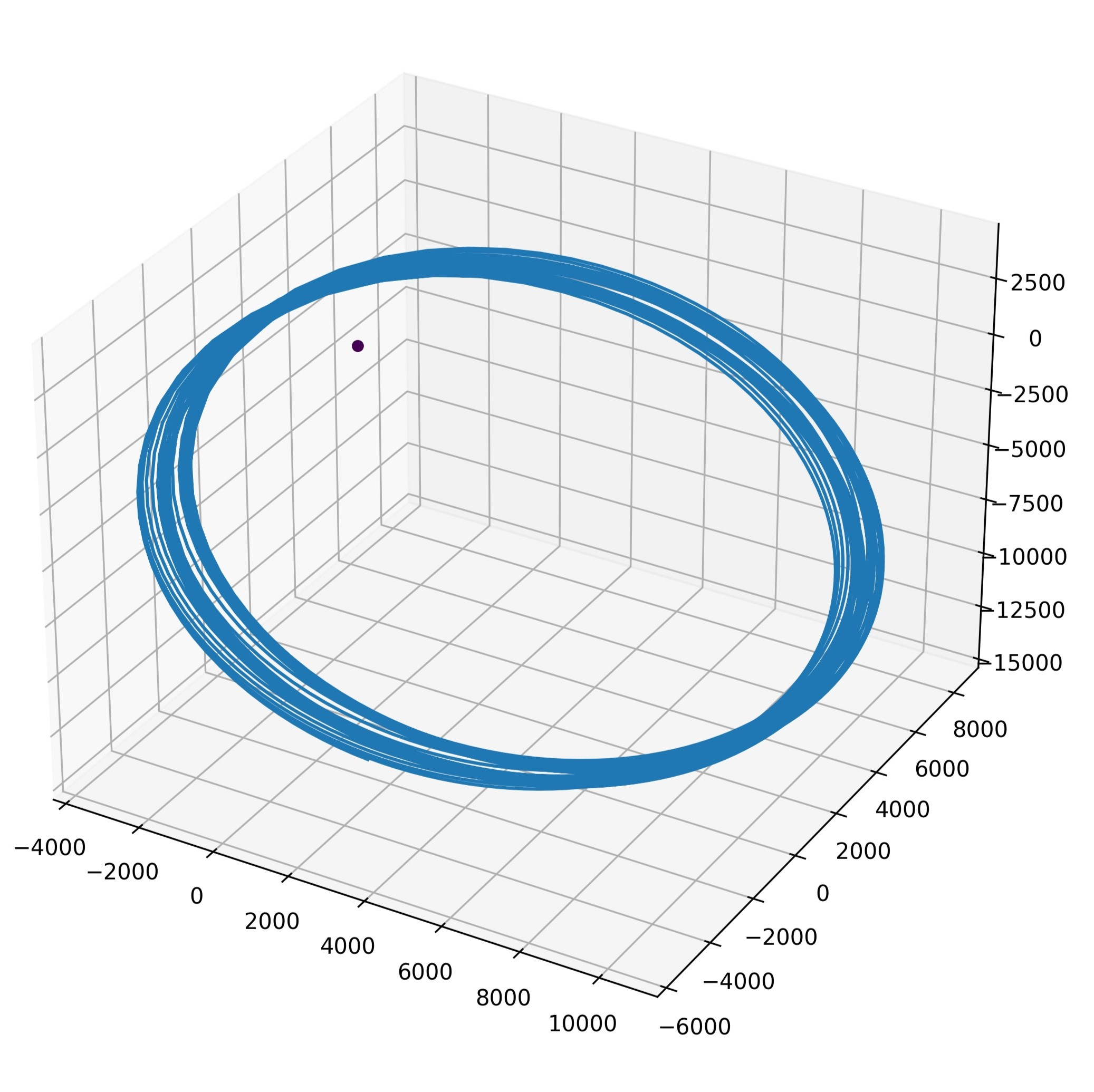}
\caption{Spacecraft position from June 01, 2026 to June 30, 2026, with the center of Moon at the origin. The violet dot represents the location of the simulated LLRR at the North Pole.}
\label{fig:fitem1} 
\end{figure}

This ranging experiment is carried out for a period of 1 year from the initialization period, i.e. until 2027 June 1 00:00:00 TDB. \colorbox{white}{The motion of the spacecraft is integrated using GODOT. It is constrained by several} components including ephemeris, gravity, etc. During this period, we sampled or had ranges at 368 instances. These 368 ranges correspond to times when the proposed North Pole-spacecraft range is the shortest which also implies that the time taken by spacecraft for one revolution around Moon is close to 1 day. The position of the spacecraft from June 01, 2026 to June 30, 2026 is shown in Figure \ref{fig:fitem1}.

Once again, we use the ESA/ESOC flight dynamics software GODOT for integrating the orbit of the spacecraft and estimating the spacecraft-North Pole range. The integration of the spacecraft orbit is done in the ICRF centered on the Moon. Later the vectors to the spacecraft and the North Pole are resolved to compute the spacecraft-North Pole range using the procedure descibed in Sections  \ref{sec:PA} and \ref{sec:sim_lrr}. A simplification is done here in not applying the PA transformation also to the gravity field coefficients. 

Since the spacecraft is operational from June 01, 2026 we extended the existing LLR observations from January 01, 2022 to December 31, 2027 for a period of 6 years with a statistical sampling similar to that from January 01, 2015 to December 31, 2020. As discussed earlier, one observation was simulated to North Pole LLRR for every 3 observations to Apollo 15 reflector and this accounted for 1864 observations between the beginning of 2014 and the end of 2027.

The covariance matrix is updated with the updated partial derivatives including the additional observations (to existing LLRRs and simulated LLRR at the North Pole) and ranges (between the LLRR at the North Pole and the spacecraft). This process involves having four assumptions for range uncertainty (RUAS 1, RUAS 2, RUAS 3 and RUAS 4) for the simulated North Pole LLRR as discussed in the section \ref{sec:sim_lrr}. In addition to this, it is necessary to have an uncertainty estimate for the range between North Pole LLRR and the spacecraft. Several simulations were carried out with different uncertainties from 10 m to 1 cm. Driven by our aim to improve uncertainty, it was found that noticeable improvement was found only when the range uncertainty is as low as 1 cm. 

\begin{figure}
\centering
\includegraphics[scale=0.43]{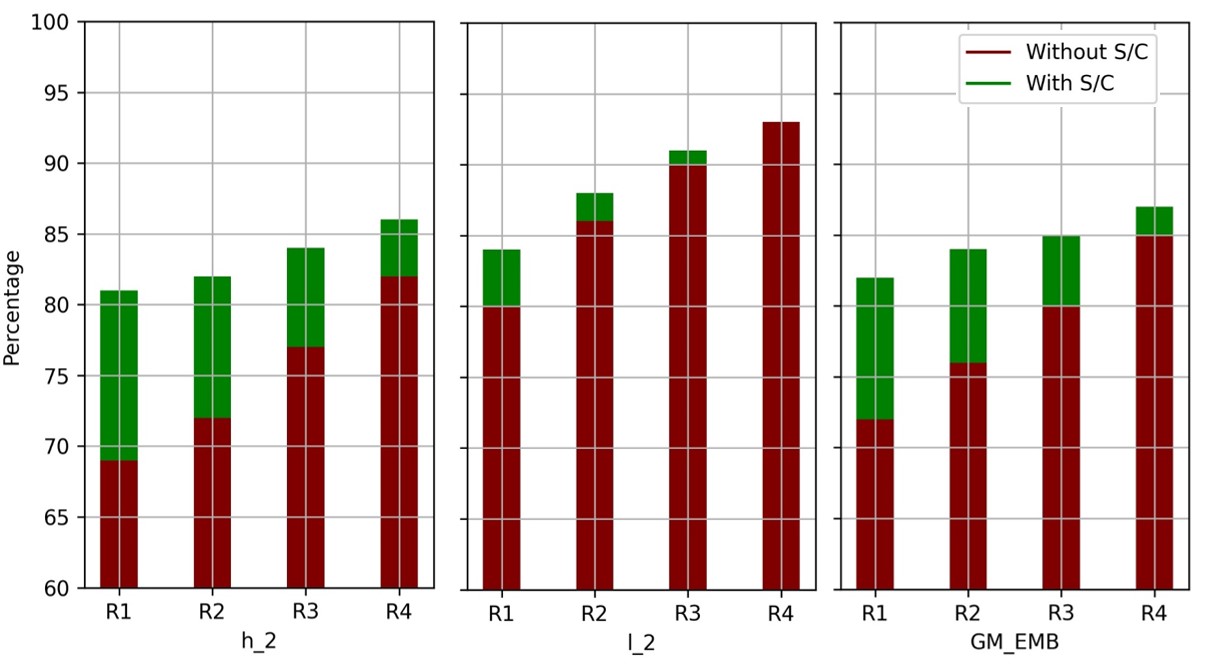}\includegraphics[scale=0.43]{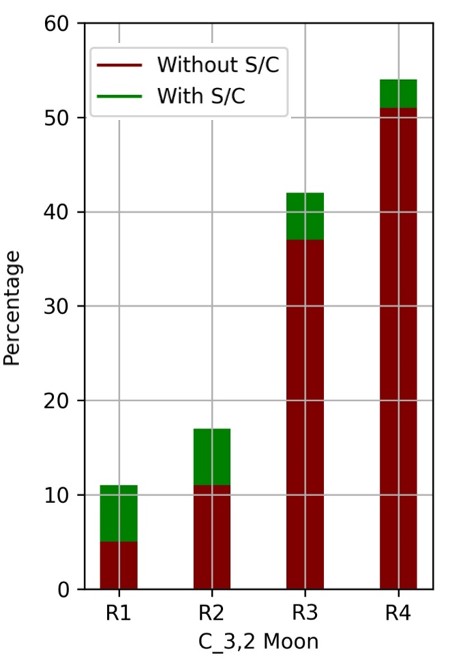}
\caption{ Improvement in $\%$ for h\textsubscript{2}, l\textsubscript{2}, GM\textsubscript{EMB} and C32 on the addition of spacecraft with 1-cm accurate range measurement between the s/c and the North Pole LLRR,  along with the LLRR range uncertainty assumption sets.}
\label{fig:hlg_sc}
\end{figure}

\begin{figure}
\centering
\includegraphics[scale=0.55]{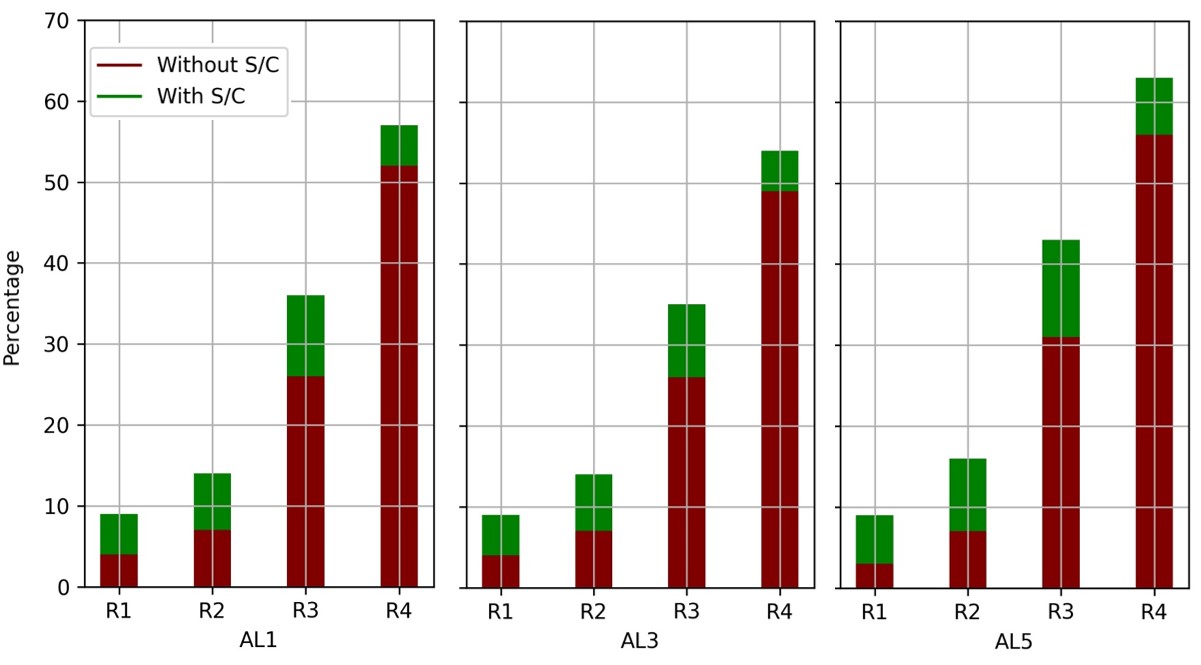}\\
\includegraphics[scale=0.53]{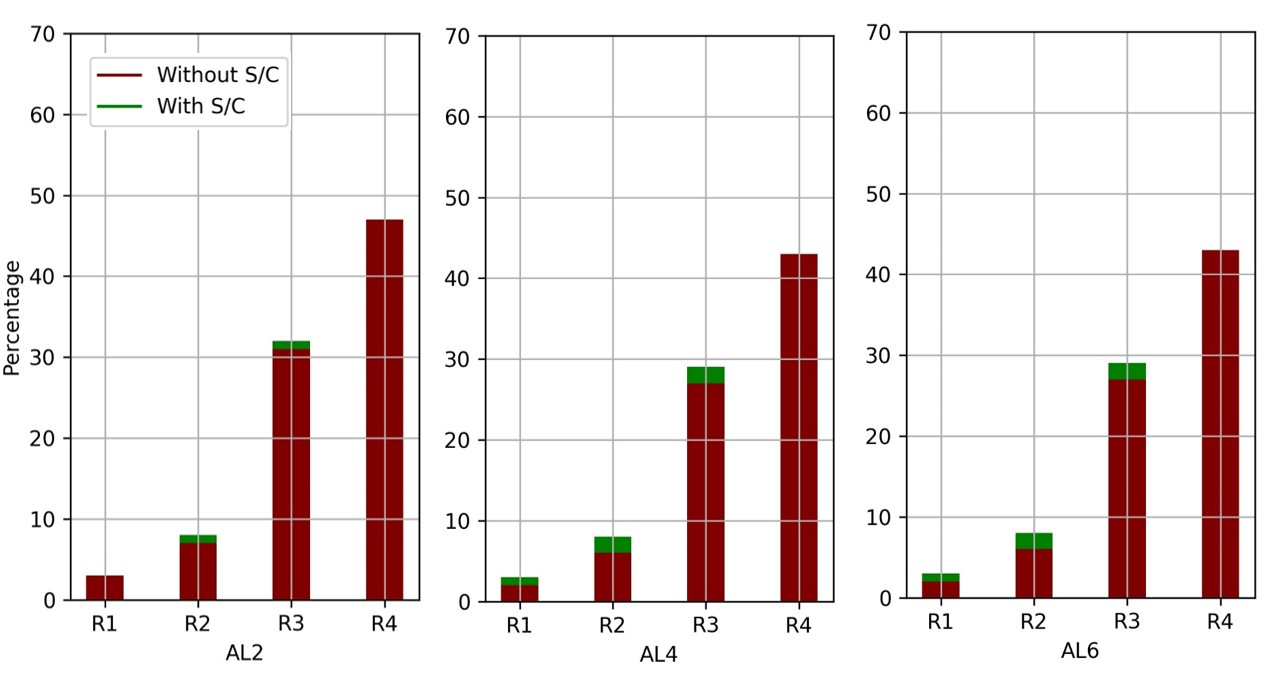}
\caption{ Improvement in $\%$ in libration angles (AL) on the addition of spacecraft with 1-cm accurate range measurement between the s/c and the North Pole LLRR,  along with the LLRR range uncertainty assumption sets. }
\label{fig:AL_sc}
\end{figure}

Not all the parameters have their accuracies improved by the use of altimetric measurements in addition to the North Pole LLRR. One Figure \ref{fig:hlg_sc} and \ref{fig:AL_sc}, are plotted for only parameters sensitive to MoonLight altimetry, the improvements in $\%$ induced by the use of North Pole LLRR in one hand (without s/c) and the use of North Pole LLRR and MoonLight altimetry (with s/c ) in the other hand for for the 4 scenarii of LLR accuracies considered in Section \ref{sec:sim_lrr} and in considering an accuracy of 1 cm for the MoonLight altimetry.

As one can see on Figure \ref{fig:hlg_sc}, the tidal Love number corresponding to horizonal solid body tide, l\textsubscript{2} benefits the most by the addition of the ranges from Earth to North Pole but it is h\textsubscript{2} that sees its accuracy improved the most by the addition of the ranges between spacecraft and the LLRR at the North Pole for all the RUAS. There is a 13 \% improvement in the corresponding uncertainty with RUAS 1 and LLRR-spacecraft range uncertainty at 1 cm. For parameters like tidal Love number corresponding to horizontal solid body tide, l\textsubscript{2} and the gravitational mass of the Earth-Moon barycenter, GM\textsubscript{EMB} the difference in improvement with and without s/c, is 4\% and 10\% respectively with RUAS 1. But for these two parameters, the improvement decreases significantly when the accuracies of the LRR increases for the RUAS 2, RUAS 3 and RAUS 4. By improving significantly the accuracy of the LLR, most of the information on l\textsubscript{2} and the gravitational mass of the Earth-Moon barycenter, GM\textsubscript{EMB} is brought by LRR and the addition of the s/c ranges with an accauracy of 1 cm does not help for improving their determinations.

\colorbox{white}{Similarly, the improvement in the uncertainty of libration angles} with spacecraft in multiple spacecraft-LLRR range uncertainty scenarii is shown in figure \ref{fig:AL_sc}. AL1, AL3 and AL5 see a good sensitivity to s/c ranges whatever is the RAUS. For AL2, AL4 and AL6, the information brought by MoonLight altimetry is marginal for all RAUS.
Finally one can notice the sensitivity of the degree 3 order 2 C coefficient of the Moon gravity field which also seems to gain in accuracy thanks to the MoonLight 1-cm altimetry.

In conclusion, the addition of ranges from LLRR-spacecraft in the MoonLight constellation is indeed improving the uncertainty estimate for some parameters (mainly the Love numbers, the GM\textsubscript{EMB} and the C32) in a scenario where the range between the s/c and the North Pole reflector can be done with at least a 1-cm accuracy. The impact on the PA angles is marginal but non negligible. The size of the improvement may also be dependent on several parameters especially the uncertainty in the LLRR-spacecraft range (RUAS) as discussed earlier.

\section*{Acknowledgements}
The authors wish to thank the participants of the ATLAS consortium, to ESA members P. Giordano, R. Swinden and  J. Ventura-Traveset as well as J. Laskar and M. Gastineau for the development of the INPOP planetary ephemerides

\bibliographystyle{spbasic}
\bibliography{global}   


\end{document}